\documentclass[a4paper,11pt]{article}
\usepackage{amsmath}
\usepackage{amssymb}
\usepackage{bm}
\usepackage{graphicx}
\usepackage{subcaption}
\usepackage{float}

\usepackage{authblk} 

\newcommand{\myrot}{\textrm{curl}}
\newcommand{\mydiv}{\textrm{div}}
\newcommand{\mygrad}{\textrm{grad}}

\def\onedot{$\mathsurround0pt\ldotp$}

\def\cddot{
  \mathbin{\vcenter{\baselineskip.67ex
    \hbox{\onedot}\hbox{\onedot}}%
  }}%

\title
  {Source identification in the self-potential method and its connection to Stokes type systems}
  
\author[1,2]{M. Malovichko}
\author[1,2]{N. Yavich}
\affil[1]{CDISE, Skolkovo Institute of Science and Technology, Moscow, Russia}
\affil[2]{Applied Computational Geophysics Lab, Moscow Institute of Physics and Technology, Dolgoprudny, Russia}

\begin{document}

\label{firstpage}

\maketitle

\begin{abstract}
This paper develops a novel approach to the problem of source current identification for the diffusion equation in connection with geophysical self-potential measurements.
The problem is split into two subproblems: (a) the scalar source identification, and (b) solution of the divergence equation. 
For subproblem (a), we design an algorithm for reconstructing the scalar source function, which 
does not require solving the Fredholm integral equation of the first kind.
Instead, the problem is reformulated as a linear operator equation, which is solved by a projection 
method.
The dimension of the subspace, in which the source function is sought, is independent of the dimension of the forward problem, 
leading to reduction of the size of the inverse problem.
Numerical experiments with exact and noisy data are presented.
For subproblem (b), the divergence equation was posed as a minimization problem, which, 
by means of Lagrangian formalism, was reduced to a system of partial differential equation of Stokes type with a unique solution.
To demonstrate how this framework can be used in a practical application, 
we implemented the algorithm in the two-dimensional physical space, using a finite-different discretization on staggered grids.

\end{abstract}

\section{Introduction}

The problem of source identification has always been a major focus in exploration geophysics.
We are interested in such a problem arising in self-potential measurements, especially, 
in connection with imaging of the seepage pathways and distribution of electrochemical potential.
For overview of the physical phenomena and recent developments we refer to \cite{Castermant2008,Revil2013,Rittgers2013,Bernabe2015,Guarracino2018}.
Mathematically, the problem is formulated as the (scalar or vector) source inverse problem of the diffusion equation.

Let $\Omega$ be a bounded domain in $\mathbb{R}^n$, $n=2,3$ with boundary $\Gamma$.
Without loss of generality, we may think of $\Omega$ as a parallelepiped domain in $\mathbb{R}^3$ (or $\mathbb{R}^2$), 
$\Gamma=\Gamma_1 \cup \Gamma_2$, where $\Gamma_2$ represents the top face of the modeling domain (the air-ground interface),
with $\Gamma_1$ being  the other faces.
The forward problem of the diffusion equation is formulated as follows:
\begin{equation}\label{eq:forward_problem}
	\begin{aligned}
		-\mydiv (\sigma \, \mygrad u ) =f \quad \text{in} \quad \Omega , \\
		u=u_0 \quad \text{on} \quad \Gamma_1, \quad
		\frac{ \partial{u} }{ \partial \bm \nu}=q_0 \quad \text{on} \quad \Gamma_2.
	\end{aligned}
\end{equation}
Here $u$ is the scalar electric potential, $\sigma$ is the electric conductivity, vector $\bm \nu$ is a unit outward normal.
The right-hand side, $f$, represents the electrical charge, generated by external currents, thus it has sign.
In geophysical applications, the values of $u_0$ are usually set to the values of electric potential 
computed for a layered background conductivity, and $q_0$ is set to zero.
Problem \eqref{eq:forward_problem} is well-posed 
(that is, it has a unique solution and stable).
In this paper we study two inverse problems, associated with forward problem \eqref{eq:forward_problem}.
\textit{The scalar source identification} problem reads:
\begin{equation}\label{eq:inverse_problem}
	\begin{aligned}
		-\mydiv (\sigma \, \mygrad u ) =f \quad \text{in} \quad \Omega,\\
		u=u_0 \quad \text{on } \Gamma_1, \quad
		\frac{ \partial{u} }{ \partial \bm \nu}=q_0 \quad \text{on} \quad \Gamma_2,\\
		\text{determine } f \text{ provided } \\
		Q(u)=u_2.
	\end{aligned}
\end{equation}
Here operator $Q$ is an observation operator.
Problem \eqref{eq:inverse_problem} is ill-posed. 
More precisely, it is not unique, 
and a solution, if exists, does not depend continuously on data $u_2$.
\textit{The current identification problem} can be formulated as follows:
\begin{equation}\label{eq:inverse_problem2}
	\begin{aligned}
		-\mydiv (\sigma \, \mygrad u ) =\mydiv(\bm j)  \quad \text{in} \quad \Omega,\\
		u=u_0 \quad \text{on } \Gamma_1, \quad
		\frac{ \partial{u} }{ \partial \bm \nu}=q_0 \quad \text{on} \quad \Gamma_2,\\
		\text{determine } \bm j \text{ provided } \\
		Q(u)=u_2,
	\end{aligned}
\end{equation}
where $\bm j$ represents the unknown distribution of current density (a vector field).
Despite looking similar, problem \eqref{eq:inverse_problem2} are substantially harder to solve than \eqref{eq:inverse_problem}.

The fundamental mathematical aspects of the scalar inverse problem \eqref{eq:inverse_problem} have been extensively 
studied, so we briefly restate a few properties important in the context of geophysical applications.
In general, the right-hand side $f$ cannot be fully restored even
if complete observations on $\Gamma$ are available.
There are conditions under which problem \eqref{eq:inverse_problem} has a unique solution \cite{Isakov,Prilepko},
but they are too restrictive for geophysical applications.
We give one such formulation, which, probably, be of most practical interest. 
Let us assume, that the potential satisfies Poisson's equation with Dirichlet boundary conditions:
\begin{equation}\label{eq:poisson}
	\begin{aligned}
		-\Delta u = f \quad \text{in } \Omega,\\
		u=u_0 \text{ on } \Gamma.\\
	\end{aligned}
\end{equation}
Let us further assume that the right-hand side is harmonic, $\Delta f=0$. 
Under these assumptions $f$ can be uniquely restored by its exterior potential \cite[Theorem 3.7.3]{Prilepko}.
Applying the Laplacian to \eqref{eq:poisson} 
we obtain the fourth-order PDE, which we supplement with two boundary conditions to get the following forward problem:
\begin{equation}\label{eq:fourh_order_pde}
	\begin{aligned}
		-\Delta^2 w = 0 \quad \text{in } \Omega,\\
		w=u_0 \text{ on } \Gamma,\quad
		\frac{\partial w}{\partial \bm \nu} =\psi \text{ on } \Gamma.
	\end{aligned}
\end{equation}
Solving \eqref{eq:fourh_order_pde} for $w$ we then determine the right-hand side of \eqref{eq:poisson} by setting $f=-\Delta w$.

The source function $f$ is easier to estimate when it comes in the factorized form, one part of which is known.
For example, if $\Omega \subset \mathbb{R}^2$, and the source is known to satisfy $f(x,y)=g(x)h(y)$ with, say, $h(y)$ being given, 
then the problem \eqref{eq:inverse_problem} is easier to solve.
In this case, under rather mild conditions, the source can be fully accessed from boundary measurements.
For further discussion and a particular example we refer to \cite{ElBadia1998}.
Even if these conditions do not hold, the regularized solution of the inverse problem better resolves the true
source function if one factor of it is known a priori.

There are exists a number of approaches to the problem, proposed in various fields, for example, \cite{ElBadia2000,Nara2003,Ling2005,Majeed2017}.
Many of them are not applicable to geophysics due to restrictive assumptions,
aiming to establish an analytical relationship between components of the right-hand side and measured data.
In geophysical literature, e.g. \cite{Minsley2007,Castermant2008}, the problem of scalar source identification is usually reduced to 
the Fredholm integral equation of the first kind with singular kernel:
\begin{equation}\label{eq:fredholm_scalar}
	\int_{\Omega} g(\bm r,\bm r') f(\bm r') d V' = u_2,
\end{equation}
where $g$ is the scalar Green function. 
This approach works reasonably well in practice, though there are a few minor caveats 
regarding to the computing of the matrix of the integral operator.
We return to this point in the next section.

The major difficulty of problem \eqref{eq:inverse_problem2} is connected to the fact that it is severely undetermined.
This point becomes obvious if we split problem \eqref{eq:inverse_problem2} into two sub-problems: 
(a) solve problem \eqref{eq:inverse_problem} for $f$, then (b) solve the divergence equation
\begin{equation}\label{eq:divergence_equation}
		\mydiv \bm j = f \quad \text{in } \Omega.
\end{equation}
Comparing to \eqref{eq:inverse_problem}, problem \eqref{eq:inverse_problem2} requires solving the divergence equation \eqref{eq:divergence_equation},
which has large null-space.

There are many algorithm to solve \eqref{eq:inverse_problem2}, for example \cite{Veen1997,Yamatani1998,ElBadia2000}.
In geophysical community an almost universally adopted strategy is to reduce problem \eqref{eq:inverse_problem2} to the Fredholm integral equation of the first kind
\cite[among many others]{Portniaguine2002,TrujilloBarreto2004,Jardani2008,Boleve2009,Ahmed2013,Rittgers2013,Ikard2014}.
The problem is expressed as
\begin{equation}\label{eq:fredholm_vector}
	\int_{\Omega} G(\bm r,\bm r') \bm j(\bm r') d V' = u_2.
\end{equation}
Here $G$ is the Green tensor.
In this formulation deficiency of \eqref{eq:fredholm_vector} may not be apparent.
Still, the integral operator in \eqref{eq:fredholm_vector}, which maps a current distribution to data, has non-empty null-space.
For a numerical solution of \eqref{eq:fredholm_vector} to make sense, it must be constructed in the visible 
subspace. 
An elegant example is provided in \cite{Magnoli1997}. 
However, it relies on a simple shape of the domain and uniform coefficients in the governing equation, 
which allows characterizing range and kernel of the integral operator analytically.
In typical geophysical settings it is not possible.
In practice, this issue is commonly tackled with Tikhonov regularization or truncated SVD,
but stable reconstruction of the current distribution remains challenging.

This paper presents a novel technique for the current identification problem, posed in the double-step form \eqref{eq:inverse_problem},\eqref{eq:divergence_equation}.
In section 2 we design an algorithm of scalar source identification, which avoids integration of a singular Fredholm kernel.
In section 3, we study a novel approach to solving the diverge equation by reducing it to a Stokes-type system.


\section{Scalar source identification}

In this section we design an algorithm for solving problem \eqref{eq:inverse_problem}.
Being an intermediate step to the current identification problem, it is important in itself, because a distribution of charges, generated by 
the ground water flows, traces the flow sources and sinks.
The solution can be computed by means of \eqref{eq:fredholm_scalar}. 
However, this formulation requires integration of the integral kernel 
in a domain of singularity to construct the matrix of the integral operator.
It can be overcome by computing the integral in the sense of principal value, 
but efforts are needed to maintain the numerical accuracy.
In what follows we employ another approach based on a projection method.

Let us specify a set of $K$ observation points in $\Omega$.
The input data, $u_2$, belong to $D = \mathbb{R}^K$.
We introduce a Hilbert space of solutions, $U$, and a Hilbert space of source functions, $F$.
The observation operator is defined as $Q : U \rightarrow D$.
We consider the following auxiliary problem: 
\begin{equation}
	\begin{aligned}
		-\mydiv (\sigma \, \mygrad v ) = 0 \quad \text{in} \quad \Omega,\\
		v=u_0 \quad \text{on } \Gamma_1,\quad
		\frac{ \partial{v} }{ \partial \bm \nu}=q_0 \quad \text{on} \quad \Gamma_2,\\
	\end{aligned}
\end{equation}
It has a unique solution. Now we consider quantity $w=u-v$, where $u$ is a solution of \eqref{eq:forward_problem}.
Obviously, $w$ is the solution to the following problem:
\begin{equation}\label{eq:final_equation}
	\begin{aligned}
		-\mydiv (\sigma \, \mygrad w ) = f \quad \text{in} \quad \Omega,\\
		w=0 \quad \text{on } \Gamma_1,\quad
		\frac{ \partial{w} }{ \partial \bm \nu}=0 \quad \text{on} \quad \Gamma_2,\\
	\end{aligned}
\end{equation}
Since $Q(w)=u_2-Q(v)$, we can regard solving problem \eqref{eq:final_equation}, 
followed by application of $Q$, as an operator that maps a given source $f\in F$ to synthetic data $Q(w)$,
$\mathcal{A}: F \rightarrow D$.
We can write down a linear operator equation:
\begin{equation}\label{eq:operator_equation}
	\mathcal{A}(f)=Q(w).
\end{equation}
The source function is expanded in an $N$-dimensional basis of some functions as follows:
\begin{equation}\label{eq:expansion}
	f(\bm r)= \sum_{n=1}^{N} a_n S_n(\bm r),
\end{equation}
where $S_n$ are basis functions of corresponding physical dimension, $a_n$ are coefficients.
A specific set of basis functions (piecewise-constant functions, wavelets, splines etc) depends on assumed properties of the solution.
For example, $f$ may be related to a solution of another partial-differential equation \cite{Boleve2009,Titov2015} and thus posses some regularity properties.

Since problem \eqref{eq:operator_equation} is linear, by means of the least-square approach we obtain the following 
system of linear equations:
\begin{equation}\label{eq:linear_system}
	\bm G \bm a = \bm b,
\end{equation}
where $\bm a =(a_1 ..a_N)^T$, 
$\bm G \in \mathbb{R}^{N\times N}$ is the Gram matrix, 
$G_{ij}=(Q(\psi_i),Q(\psi_j))_D$, 
$\psi_i=\mathcal{A}(S_i)$, 
$\bm b_i=(Q(\psi_i),Q(w))_D$,
where $(\cdot,\cdot)_D$ is the scalar product in $D$.
The condition number of $\bm G$ is likely be high, so some regularization 
is essential when solving \eqref{eq:linear_system}.

The algorithm, outlined above, avoids the difficulty, connected with integrating of singular Fredholm kernels.
The dimension of the projection subspace, $N$, is independent of the subspace of solutions of the forward problem.
This means that the size of the inverse problem can be substantially lower than that of the forward problem.


We consider the following numerical experiment.
Let $\Omega=[0,1]\times[0,1]$. 
To simplify technical details related to the forward modeling, we 
set a uniform conductivity $\sigma=1$ S/m and apply the homogeneous Dirichlet boundary conditions on the entire boundary.
Domain $\Omega$ was discretized into numerical grid 50$\times$50.
The forward problem was solved by expanding the solution to the eigenfunction of the discrete Laplacian.

\begin{figure}
	\begin {subfigure}[b]{0.49\linewidth}
		\includegraphics[width=\linewidth]{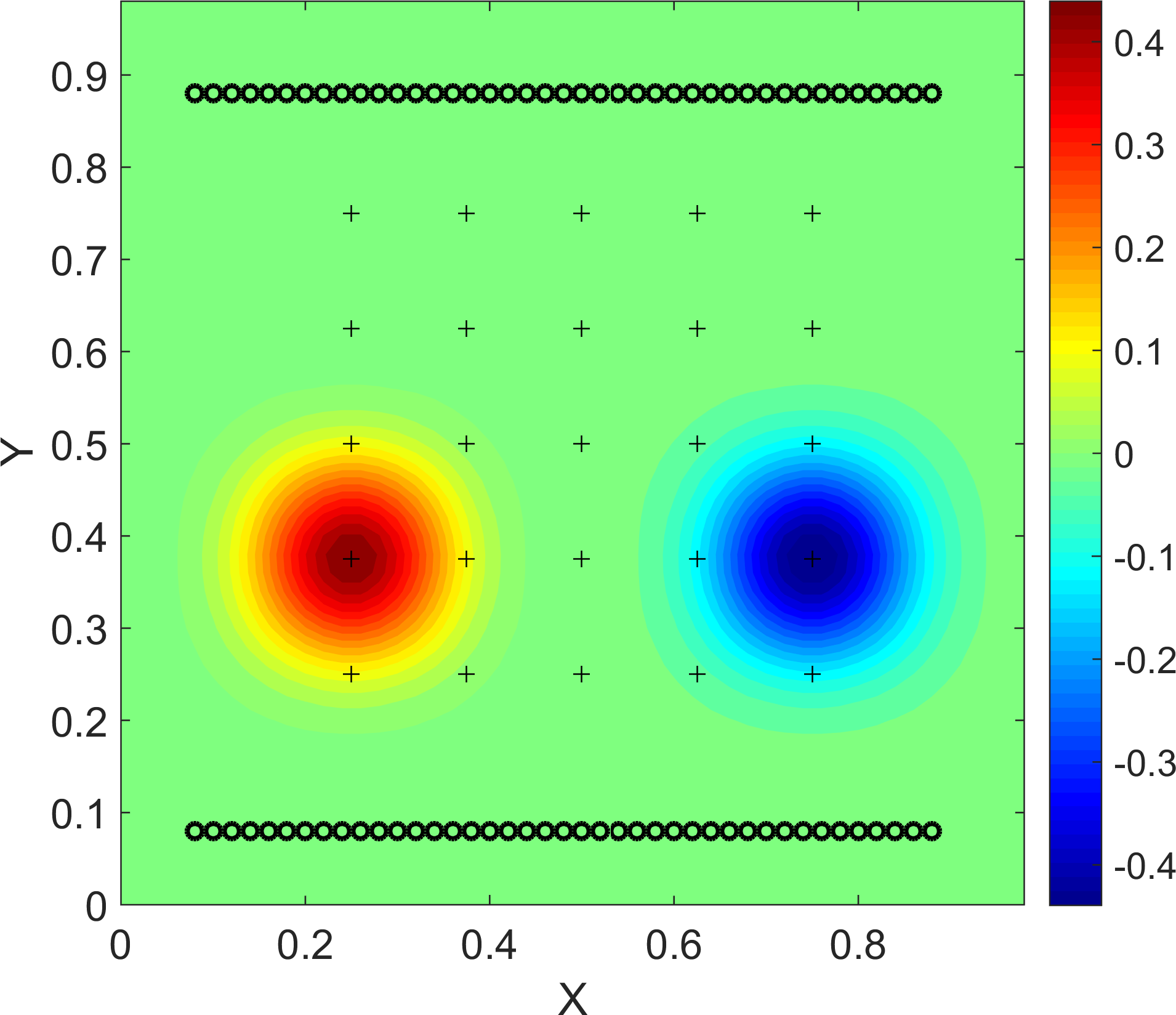}
		\caption{}
		\label{fig:rhs}
	\end{subfigure}
	\begin {subfigure}[b]{0.45\linewidth}
		\includegraphics[width=\linewidth]{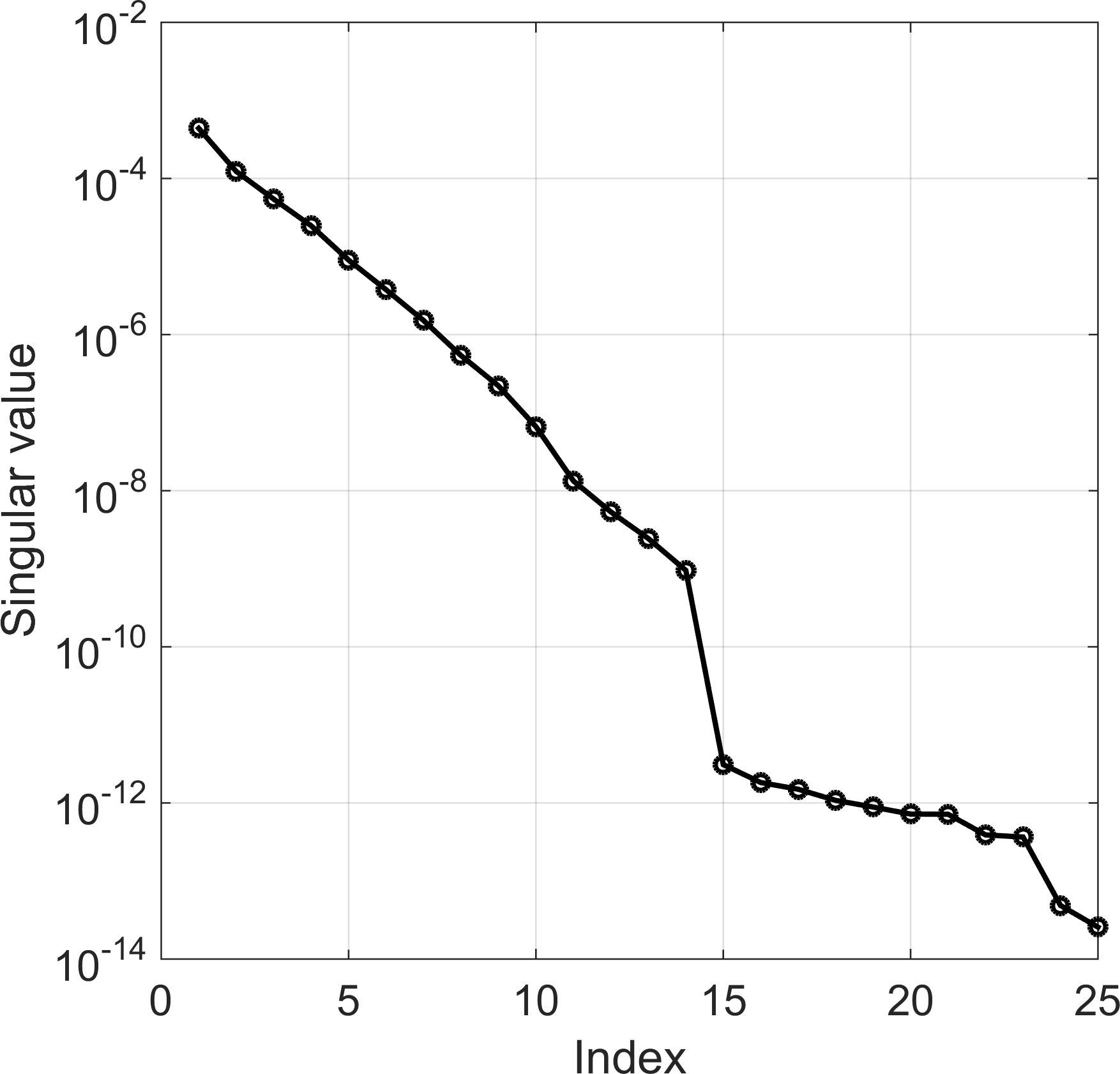}
		\caption{}
		\label{fig:SVD_spectrum}
	\end{subfigure}
	\begin {subfigure}[b]{0.49\linewidth}
		\includegraphics[width=\linewidth]{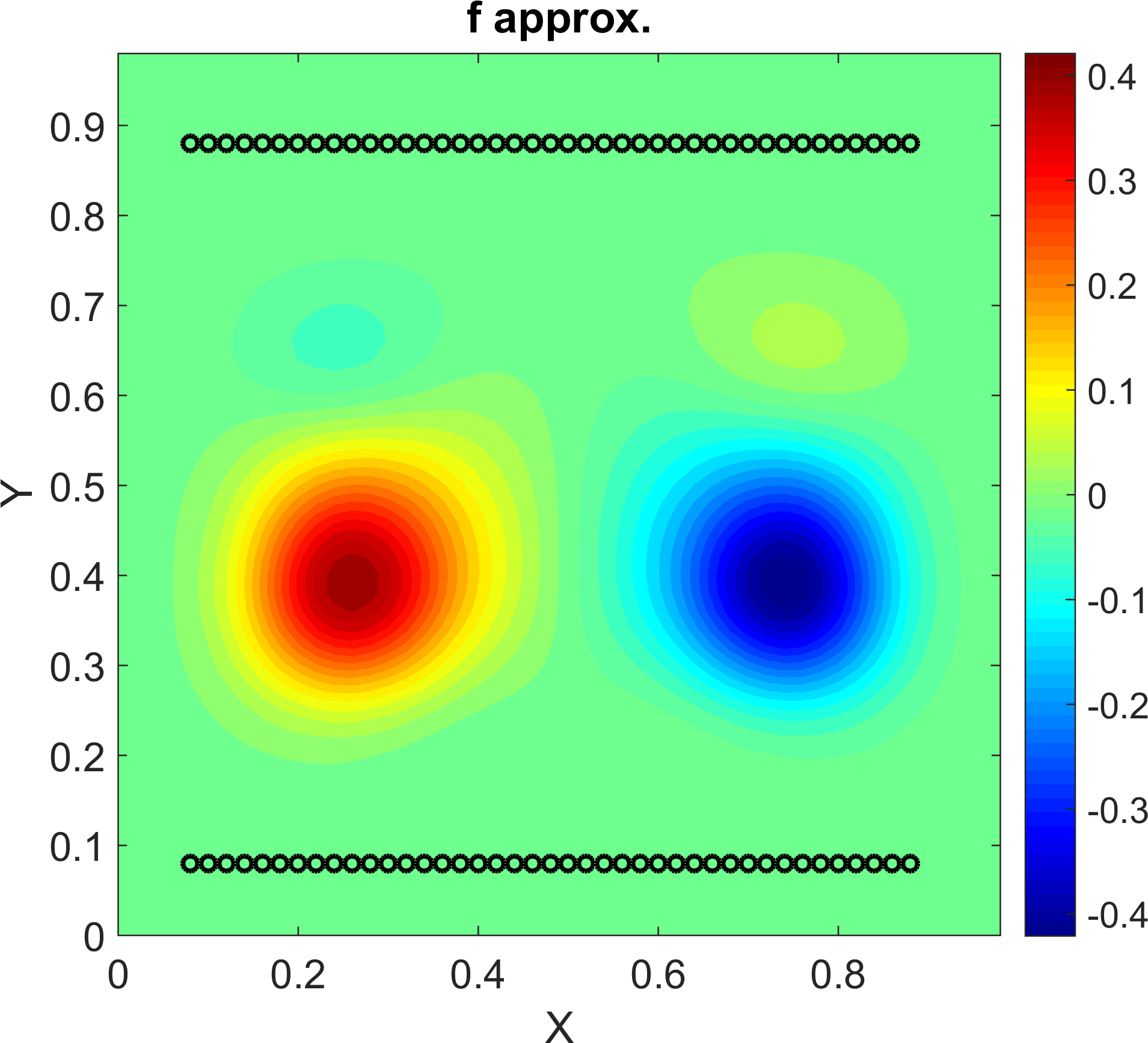}
		\caption{}
		\label{fig:exact_data_1}
	\end{subfigure}
	\begin {subfigure}[b]{0.43\linewidth}
		\includegraphics[width=\linewidth]{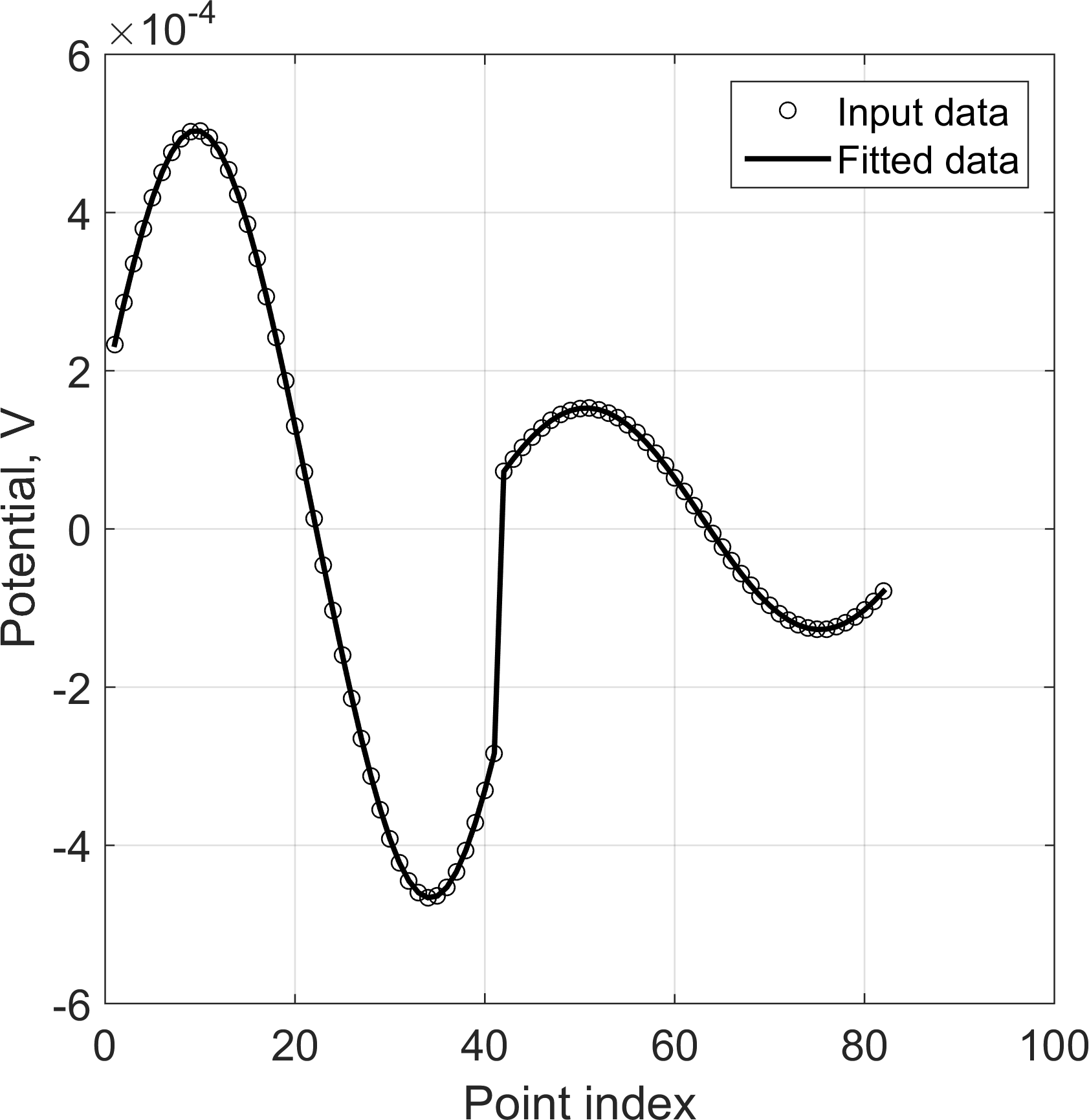}
		\caption{}
		\label{fig:exact_data_2}
	\end{subfigure}
	\begin {subfigure}[b]{0.49\linewidth}
		\includegraphics[width=\linewidth]{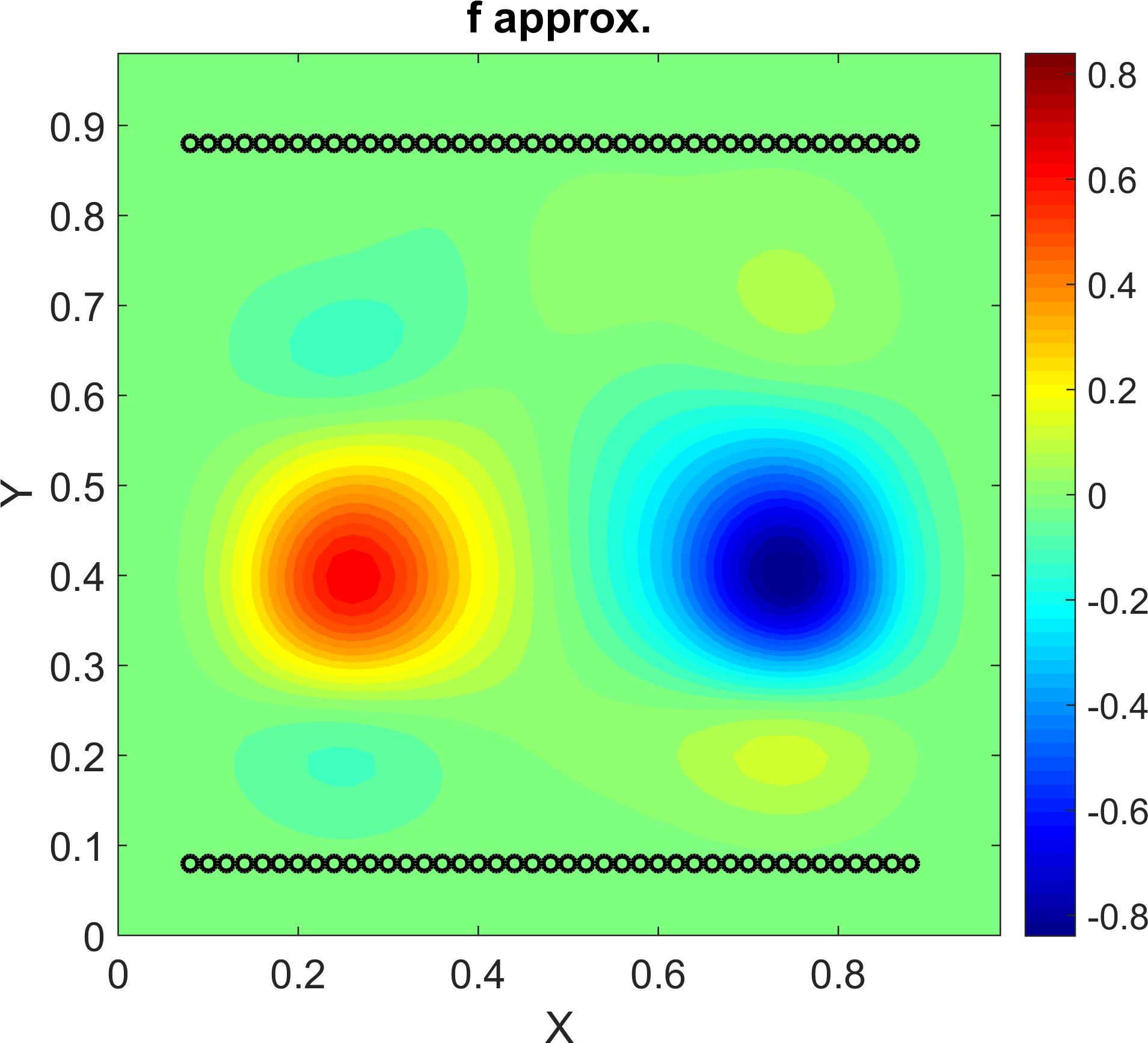}
		\caption{}
		\label{fig:noisy_data_1}
	\end{subfigure}
	\begin {subfigure}[b]{0.43\linewidth}
		\includegraphics[width=\linewidth]{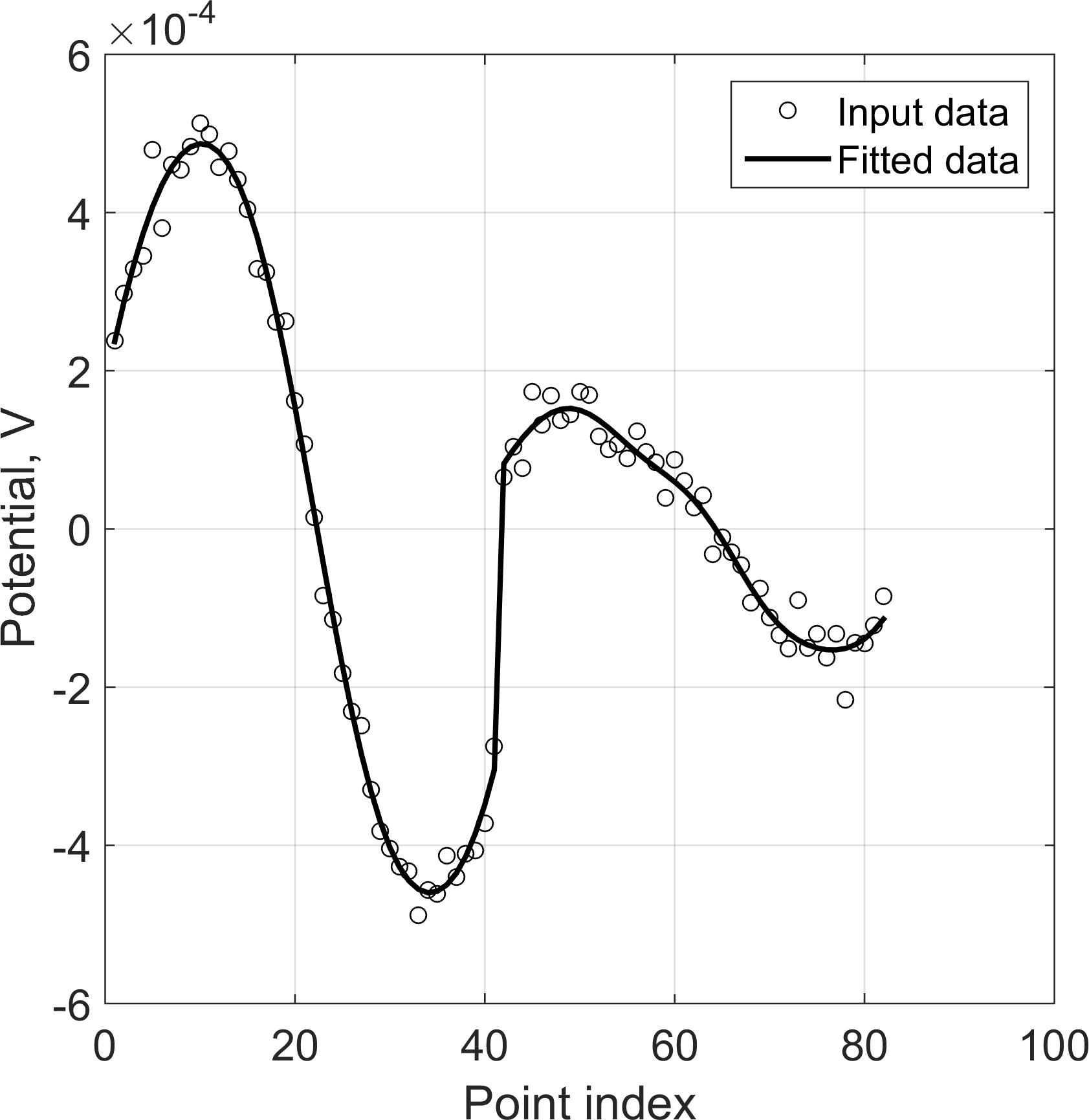}
		\caption{}
		\label{fig:noisy_data_2}
	\end{subfigure}		
	\caption{
		(a) The true source function (shown in color) and centers of splines (crosses).
		(b) Singular spectrum of the Gram matrix.
		(c) Reconstructed source function, exact data
		(d) Comparison of input and predicted measurements, exact data.		
		(e) Reconstructed source function, noisy data.
		(f) Comparison of input and predicted measurements, noisy data.		
	}
	\label{fig:numerical_experiment_1}
\end{figure}

We will seek the source term in form of linear combination of B-splines, 
thus assuming that it is of class $C^2(\Omega)$.
We introduce a rectangular mesh $I \times J$ with a step $d$ on which the two-dimensional cardinal cubic B-splines are defined:
\begin{equation}
	\begin{aligned}
		S_{ij}(x,y)=S \left( \frac{x-x_i}{d} -2\right) 
					S \left( \frac{y-y_j}{d} -2\right), \\
		S(x) = 
		\begin{cases}
			x^3/6                  ,& 0 \leq x < 1,   \\
			(-3x^3+12x^2-12x+4)/6  ,& 1 \leq x < 2,   \\
			(3x^3-24x^2+60x-44)/6  ,& 2 \leq x < 3,   \\
			(-3x^3+12x^2-48x+64)/6 ,& 3 \leq x < 4,   \\
			0                      .& \text{otherwise.}
		\end{cases}
	\end{aligned}
\end{equation}
Thus $N=I\times J$ and the source is expanded as follows:
\begin{equation}
	f(x,y)= \sum_{i=1}^{I} \sum_{j=1}^{J} a_{ij} S_{ij}(x,y).
\end{equation}
We set $I=J=5$ with $d=0.125$.
The right-hand side was set to a sum of two B-splines of different signs as shown in (Fig.~\ref{fig:rhs}).
Thus, the true source function had only two non-zero coefficients in \eqref{eq:expansion}: 1 and -1.
The system of linear equation was solved by computing its Moore-Penrose inverse.

The measurements were taken at 82 points along two lines, $y=0.1$ and $y=0.9$.
The spectrum of singular values of matrix $\bm G$ is given in Fig.~\ref{fig:SVD_spectrum}.
There is a notable gap after the first 14 singular values, so value $10^{-10}$ was used as the threshold.
The solution of the inverse problem and comparison between the measured and predicted data are shown in Fig.~\ref{fig:numerical_experiment_1}c,d.
We then contaminated the data with Gaussian noise of zero mean and standard deviation equals $3 \times 10^{-5}$.
Results and the data fit are presented in Fig.~\ref{fig:numerical_experiment_1}e,f.
We observed good data fit in all cases and decent similarity of the reconstructed source function, as compared to the true one.


\section{Current identification}

Let us assume that the divergence is known, i.e. problem  \eqref{eq:inverse_problem} has been solved exactly.
Since solenoidal currents do not contribute to the electric potential, problem \eqref{eq:divergence_equation} admits infinitely many solutions. 
Additional information must be provided.

The standard technique is to impose condition $\myrot~\bm j=0$.
The current can be expressed as gradient of an unknown potential, $\bm j=\mygrad~h$.
It leads to the following Poisson's problem:
\begin{equation}\label{eq:poisson_current}
	\begin{aligned}
		\Delta h=f \quad \text{ in } \Omega, \\
		h=0 \quad \text{on} \quad \Gamma.\\
	\end{aligned}
\end{equation}
When $h$ is found, the source current is constructed by taking gradient of $h$.
Unfortunately, numerical experiments (not presented here) show that this approach produces poor results. 
The reason is that distributions of currents due to fluid flows have a strong solenoidal mode.

The connection between the divergence equation and the fluid dynamics has been recognized for some time, 
serving mainly as a theoretical tool \cite[and references therein]{Geissert2006}.
Recently, this relationship was exploited in \cite{Caboussat2012} to solve the divergence equation numerically.
To our knowledge, this approach can be traced back to \cite{Clement1993}.
Here we apply a similar technique to the problem of current identification.
We will seek a current distribution satisfying $\mydiv \bm j = f$ and having smoothest components among all possible distributions. 
Let us consider the following minimization problem:
\begin{equation}\label{eq:minimization_problem}
	\begin{aligned}
		\Phi(\bm j) = \frac{1}{2} \int_{\Omega} |\nabla \bm j|^2 d\Omega \xrightarrow[\bm j]{} min,\\
		\text{subject to  } \mydiv \bm j=f.
	\end{aligned}
\end{equation}
Here $|\nabla \bm j|^2 = \nabla \bm j \cddot \nabla \bm j = \sum_{i=1}^{n} \nabla j_i \cdot \nabla j_i$, 
with $\cddot$ being the double dot product defined as $A \cddot B=\sum_{i=1}^{n}\sum_{j=1}^{n} A_{ij}B_{ij}$.
We form the Lagrangian as follows:
\begin{equation}\label{eq:lagrangian}
	\mathcal{L}(\bm j, p)=\frac{1}{2} \int_{\Omega} |\nabla \bm j|^2 d\Omega + \int_{\Omega} p (f-\mydiv \bm j) d\Omega,
\end{equation}
where a real-valued scalar function $p$ is the Lagrange multiplier. 
The saddle-point solution of \eqref{eq:minimization_problem} is a pair $\{\bm j_*, p_*\}$ that satisfies 
the following necessary conditions
\begin{equation}\label{eq:necessary_conditions}
	\begin{aligned}
		\nabla_j \mathcal{L}(\bm j_*, p_*) = 0,\\
		\nabla_p \mathcal{L}(\bm j_*, p_*) = 0.
	\end{aligned}
\end{equation}
The first variation of the first term of \eqref{eq:lagrangian} with respect to $\bm j$ equals to
\begin{equation}\label{eq:variation_of_term1}
	\int_{\Omega} \nabla \bm j \cddot \nabla \bm \xi \, d\Omega 
		= -\int_{\Omega} \Delta \bm j \, \bm \xi \, d\Omega 
		+ \oint_{\Gamma} \frac{\partial \bm j}{\partial \bm \nu} \, \bm \xi \, d\Gamma,
\end{equation}
where $\bm \xi$ is the variation of $\bm j$. 
The right-hand side of \eqref{eq:variation_of_term1} follows from applying the first Green's identity to the left-hand side.
The first variation of the second term of \eqref{eq:lagrangian} equals to
\begin{equation}\label{eq:variation_of_term2}
	-\int_{\Omega} p \, \mydiv \bm \xi \, d\Omega =
		\int_{\Omega} \nabla p \cdot  \bm \xi \, d\Omega  
		- \oint_{\Gamma} p \bm \nu \cdot \bm \xi \, d\Gamma,
\end{equation}
where equality follows from applying Ostrogradsky's theorem to the left-hand side.
Combining \eqref{eq:variation_of_term1} and \eqref{eq:variation_of_term2}, taking variation of \eqref{eq:lagrangian} with respect to $p$, and using conditions \eqref{eq:necessary_conditions}, we arrive to the following system:
\begin{equation}\label{eq:euler_lagrange}
	\begin{aligned}
		-\Delta \bm j + \nabla p = 0 \quad \text{in } \Omega,\\
		-\mydiv \bm j= -f \quad \text{in } \Omega,\\
	\end{aligned}
\end{equation}
\begin{equation}\label{eq:euler_lagrange_bc1}
		\frac{\partial \bm j}{\partial \bm \nu} - \bm \nu p = 0 \quad \text{on } \Gamma.
\end{equation}
Here the first and the last equations consist of $n$ equations for corresponding current components.
System \eqref{eq:euler_lagrange},\eqref{eq:euler_lagrange_bc1} is the Euler-Lagrange system associated with \eqref{eq:minimization_problem}.
It is a Stokes-type system describing steady slow motion of a fluid, driven by sources and sinks, with zero body force.
Current $\bm $ is interpreted as the velocity field, whereas $p$ has the meaning of pressure.
For its mathematical properties we refer to \cite{Glowinski2003}.

Problem \eqref{eq:euler_lagrange},\eqref{eq:euler_lagrange_bc1} can be solved numerically, as presented in \cite{Caboussat2012}, though its solution is not unique.
However, boundary conditions \eqref{eq:euler_lagrange_bc1} admit current flow across the boundaries.
For the problem under consideration it is natural to assume zero net current in $\Omega$.
We replace conditions \eqref{eq:euler_lagrange_bc1} with homogeneous Dirichlet boundary conditions:
\begin{equation}\label{eq:euler_lagrange_dirichlet}
	\bm j=0 \quad \text{on } \Gamma.
\end{equation}
Problem $\eqref{eq:euler_lagrange}$,\eqref{eq:euler_lagrange_dirichlet} has a unique solution.
Thus, the ill-posed problem \eqref{eq:divergence_equation} was reduced to a well-posed Stokes problem.

\begin{figure}
	\begin {subfigure}[b]{0.49\linewidth}
		\includegraphics[width=\linewidth]{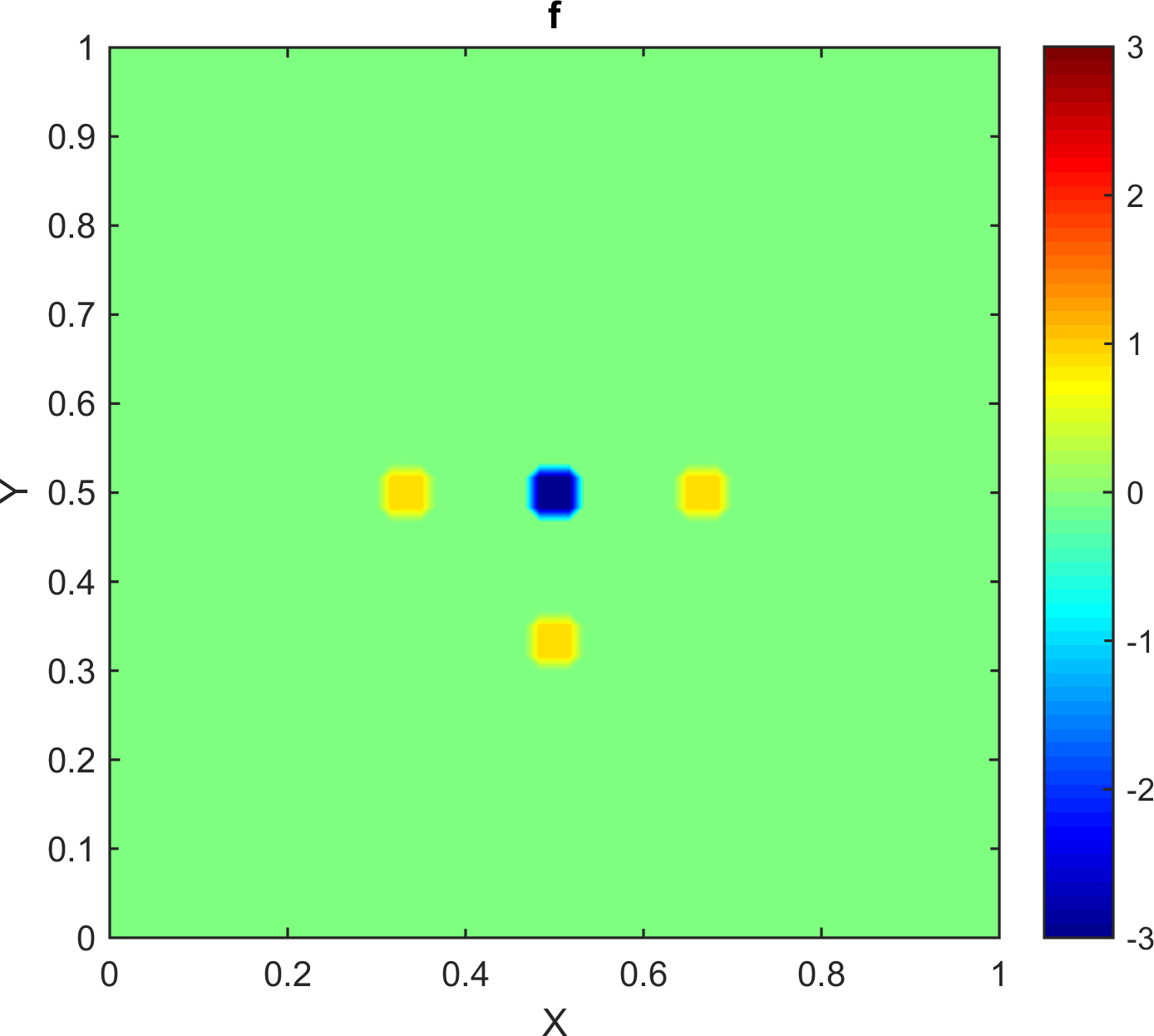}
		\caption{}
		\label{fig:divergence_rhs}
	\end{subfigure}
	\begin {subfigure}[b]{0.49\linewidth}
		\includegraphics[width=\linewidth]{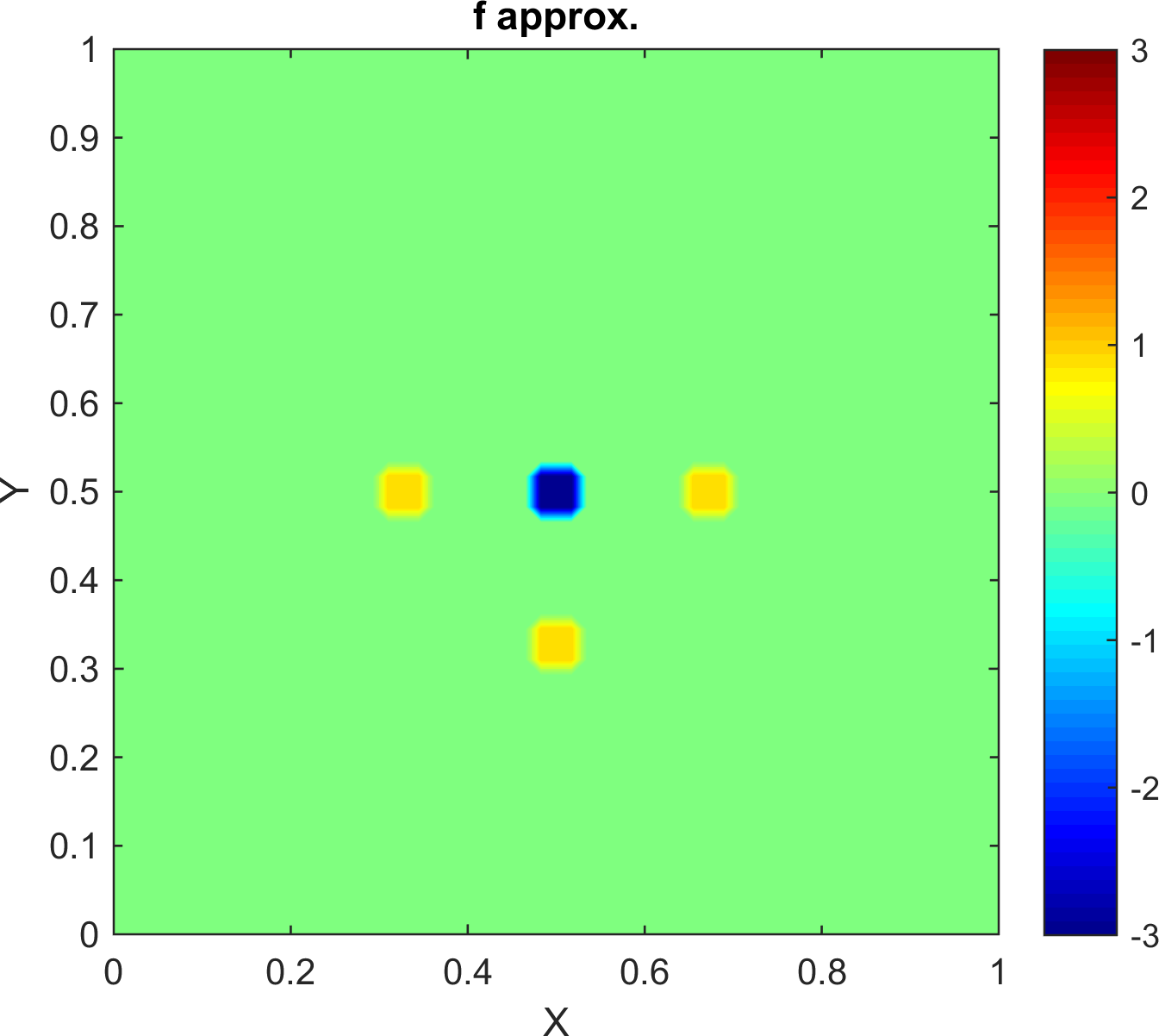}
		\caption{}
		\label{fig:divergence_rhs_predicted}
	\end{subfigure}
	\begin {subfigure}[b]{0.49\linewidth}
		\includegraphics[width=\linewidth]{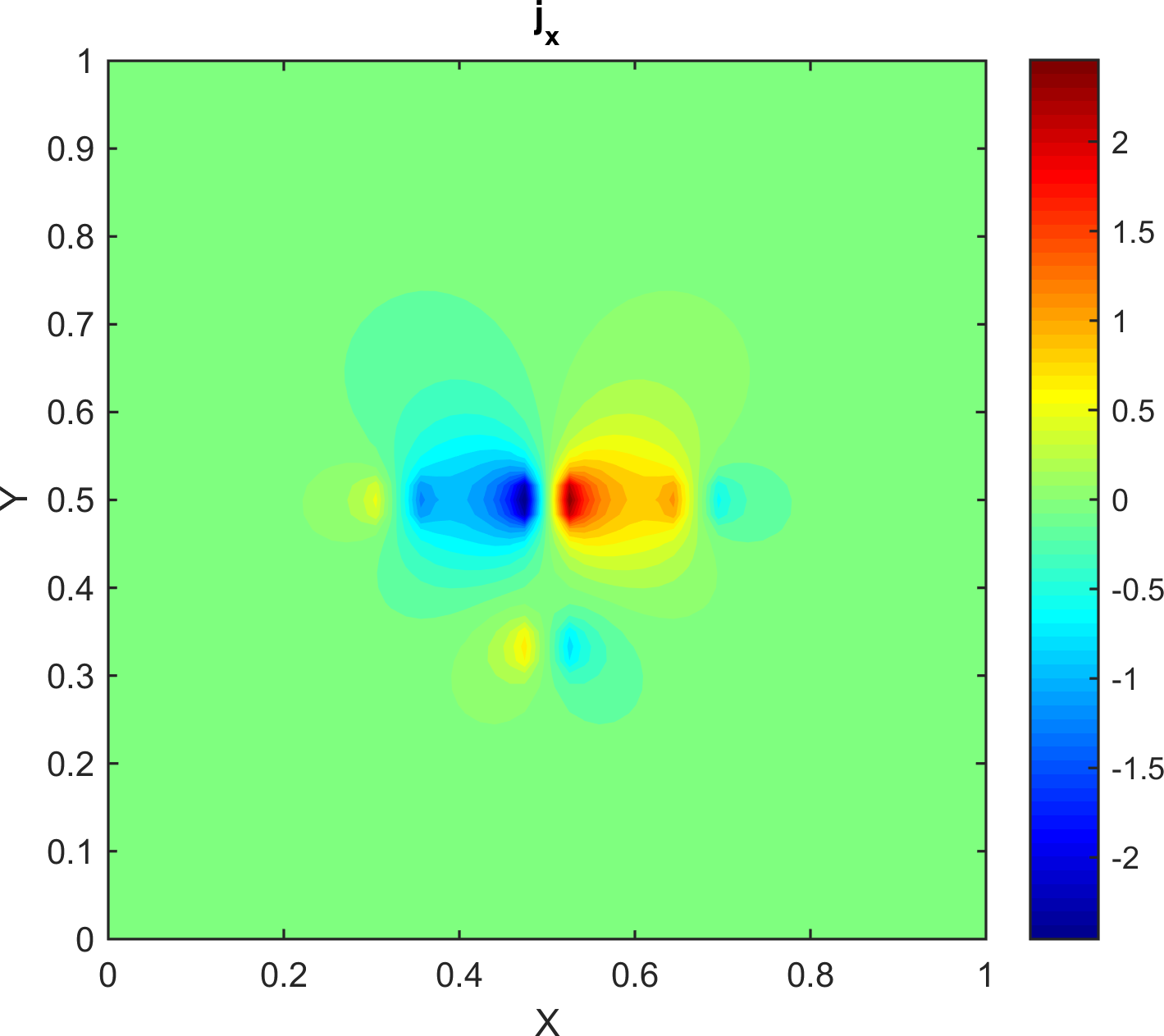}
		\caption{}
	\end{subfigure}
	\begin {subfigure}[b]{0.49\linewidth}
		\includegraphics[width=\linewidth]{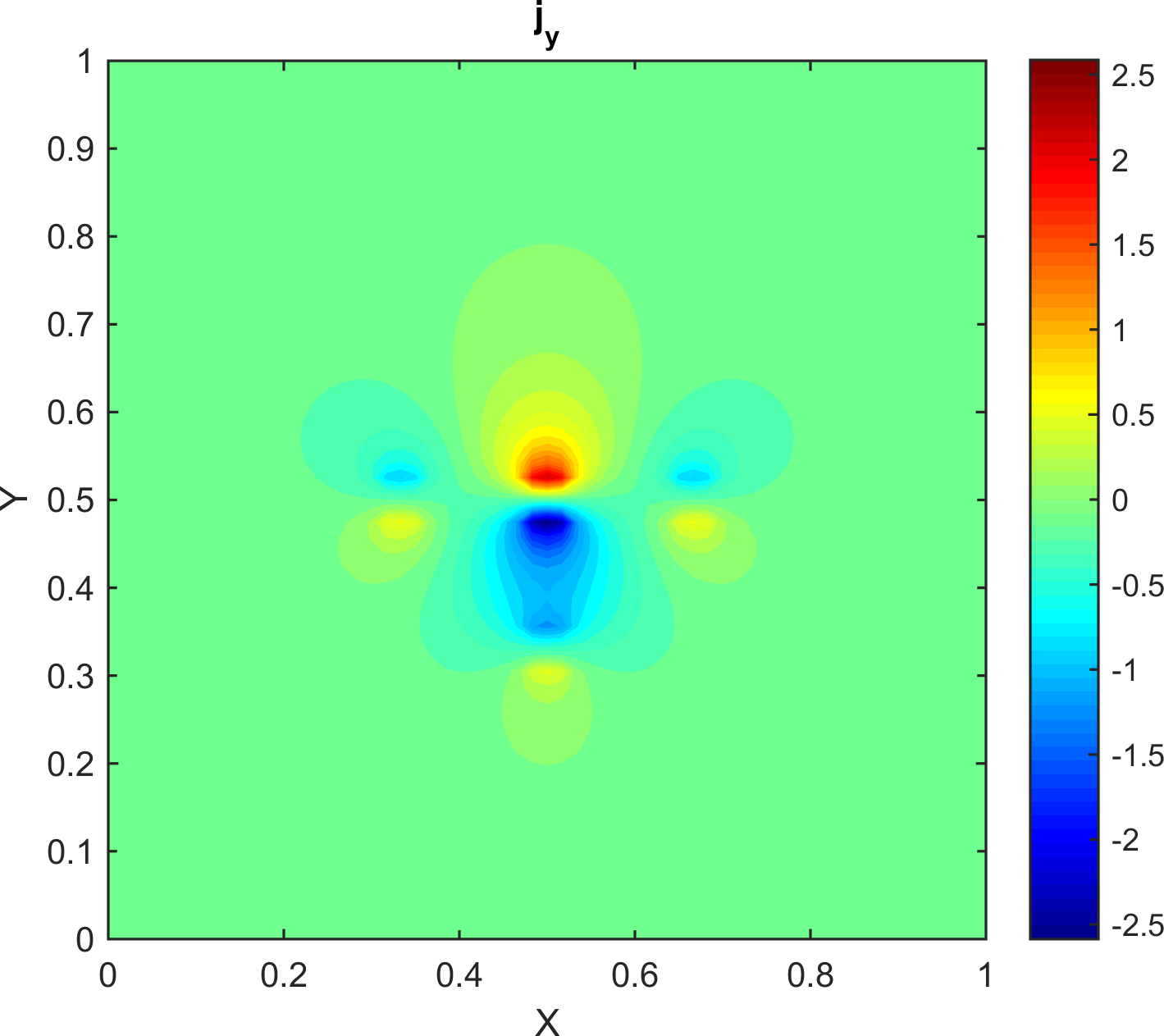}
		\caption{}
	\end{subfigure}
	\begin {subfigure}[b]{0.49\linewidth}
		\includegraphics[width=\linewidth]{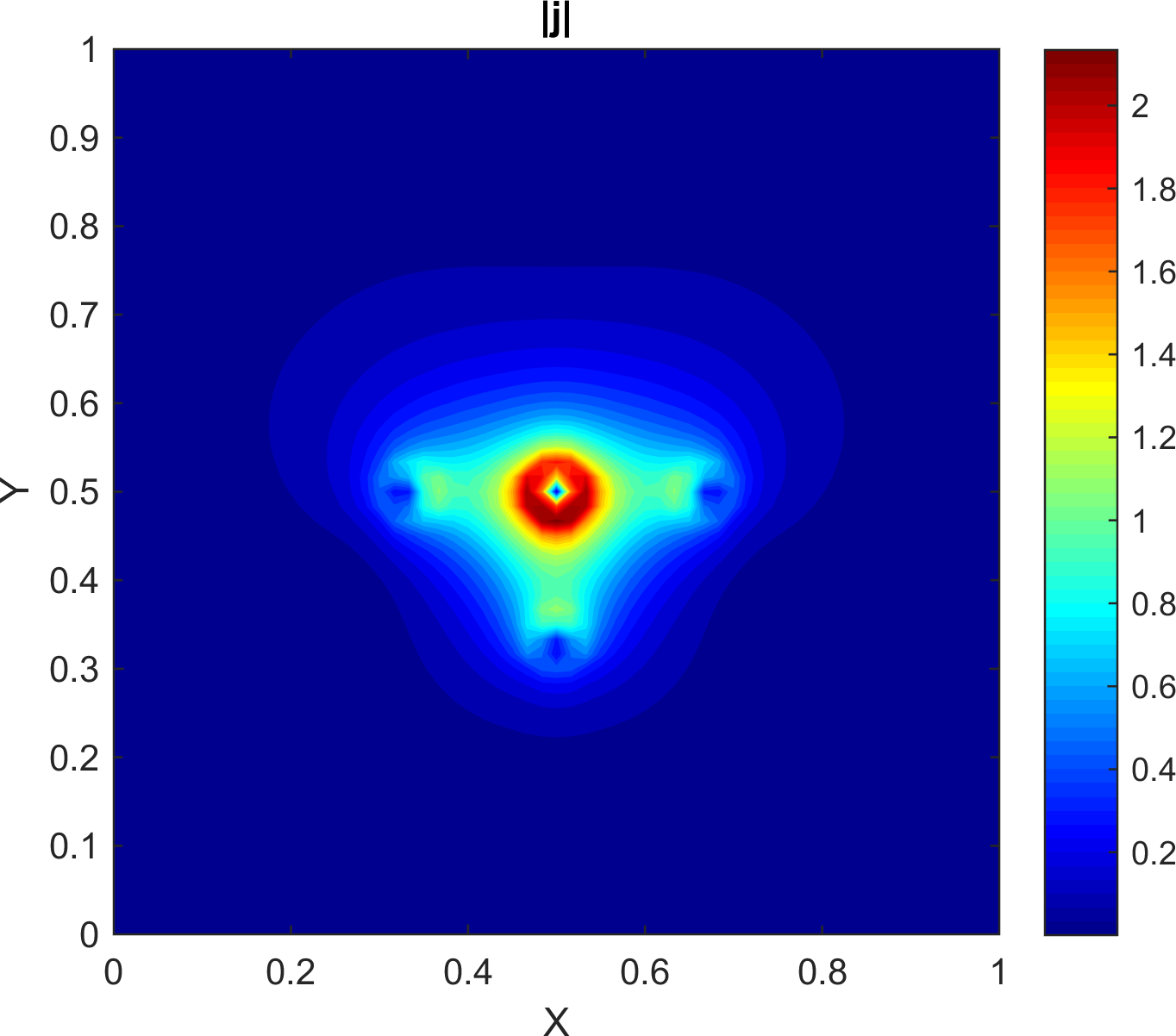}
		\caption{}
	\end{subfigure}
	\hspace{0.01\linewidth}%
	\begin {subfigure}[b]{0.42\linewidth}
		\includegraphics[width=\linewidth]{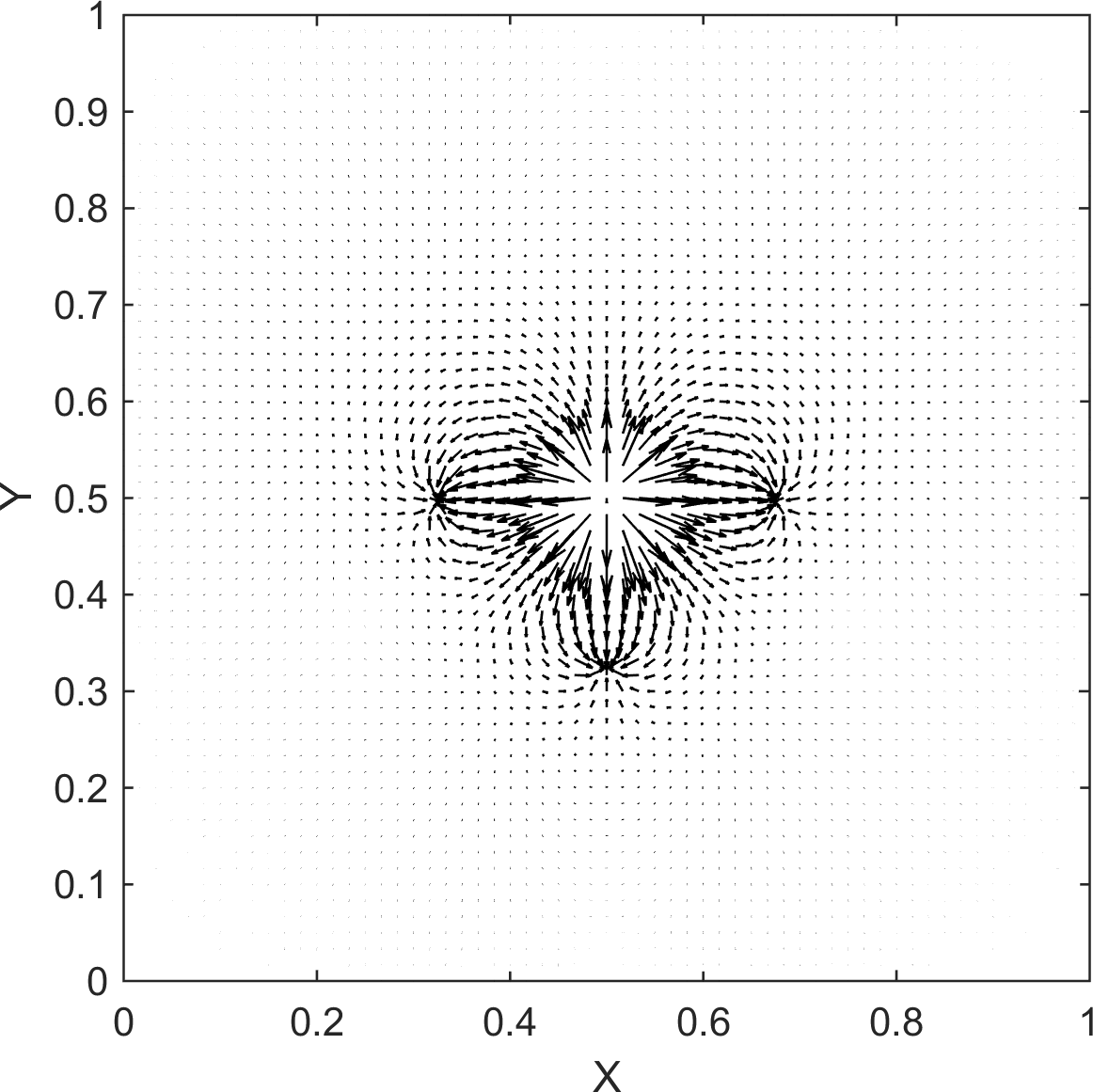}
		\caption{}
	\end{subfigure}
	\caption{
		(a) The true divergence. 
		(b) Divergence computed from the reconstructed current.
		(c) Reconstructed current, $x$-component. 
		(d) Reconstructed current, $y$-component. 
		(e) Reconstructed current, magnitude.
		(f) Reconstructed current, visualization of flows.
		}
	\label{fig:numerical_experiment_2}	
\end{figure}

To demonstrate the behavior of this method, we consider the following numerical experiment.
The divergence is given in a square $[0,1]\times[0,1]$, as presented in~Fig.\ref{fig:divergence_rhs}.
The domain was split into 61$\times$61 square cells.
Problem $\eqref{eq:euler_lagrange}$,\eqref{eq:euler_lagrange_dirichlet} reduces to the following system:
\begin{equation}\label{eq:stokes_2d}
	\begin{aligned}
		-\frac{\partial^2 j_x}{\partial x^2}-\frac{\partial^2 j_x}{\partial y^2} + \frac{\partial p}{\partial x} = 0,\\
		-\frac{\partial^2 j_y}{\partial x^2}-\frac{\partial^2 j_y}{\partial y^2} + \frac{\partial p}{\partial y} = 0,\\
		-\frac{\partial j_x}{\partial x} - \frac{\partial j_y}{\partial y} = -f,\\
		j_x(0,y)=j_x(1,y)=j_x(x,0)=j_x(x,1)=0,\\
		j_y(0,y)=j_y(1,y)=j_y(x,0)=j_y(x,1)=0.\\
	\end{aligned}
\end{equation}
To solve \eqref{eq:stokes_2d} we apply the MAC scheme \cite{Wesseling2001}, which is a finite-difference discretization on staggered grids.
The pressure is discretized in the cell centers, whereas velocity components are located on vertical ($j_x$) and horizontal ($j_y$) edges.
We used a second-order discretization inside $\Omega$. 
The boundary conditions were approximated to the first order.
It resulted to the following system of linear equations:
\begin{equation}\label{eq:stokes_linear_system}
	\left[
	\begin{array}{ccc}
	A & 0 & C \\
	0 & B & D\\
	C^T & D^T & 0
	\end{array}
	\right]
	\left[
	\begin{array}{c}
	j_x \\
	j_y \\
	p
	\end{array}
	\right]
	=
	\left[
	\begin{array}{c}
	0 \\
	0 \\
	-f
	\end{array}
	\right],
\end{equation}
where matrices $A$ and $B$ correspond to the divergence operator of components $j_x$ and $j_y$, respectively; 
matrices $C$ and $D$ correspond to the derivative operator in $x$ and $y$ directions, respectively.
The system matrix is symmetric indefinite.
There are efficient iterative solvers for systems of this type. 
We solved the system directly.
The reconstructed current is presented in Fig.~\ref{fig:numerical_experiment_2}c-f. 
It can be interpreted as a smeared image of three linear currents forming T letter. 
We emphasize that, this distribution of current fits the divergence up to numeric error (Fig.~\ref{fig:divergence_rhs_predicted}), 
and also has minimal norm of its components among all possible distributions.

\section{Conclusions}

We presented a novel framework for solving the current source identification problem
of self-potential measurements.
The framework consists of the scalar source identification followed by solution of the divergence equation.
We design an algorithm of the scalar source identification, based on posing the problem as a linear operator equation and application of a projection method. 
We also propose a method of solving the divergence equation by means of reduction to a well-posed Stokes-type system of partial-differential equations.
A few numerical experiments, given in this paper, suggest that the presented framework may have considerable potential in geophysical applications.
More research is needed to validate the efficiency of this approach to real geophysical data.
Future works will be directed to developing a practical algorithm, capable to process real measurements, in a domain with complex boundaries.


\label{lastpage}


\begin{thebibliography}{}

	\bibitem{Ahmed2013}
		Ahmed, S.A., Jardani, A., Revil, A., Dupont, J.P., 2013
		SP2DINV: A 2D forward and inverse code for streaming potential problems,
		\textit{Computers \& Geosciences}, \textbf{59}, 9-16.

	 \bibitem{Bernabe2015}
		 Bernab{\'e},Y. \& Maineult, A., 2015
		 Physics of porous media: Fluid flow through porous media, (Second Ed.). 
		 \textit{In Schubert, G. (Ed.), Treatise on geophysics (2nd ed.)}, pp. 19–41,Oxford: Elsevier.

	\bibitem{Boleve2009}
		Bol{\`e}ve, A., Revil, A., Janod,F., Mattiuzzo, J.L., \& Fry, J.-J. 2009. 
		Preferential fluid flow pathways in embankment dams imaged by self-potential tomography, 
		\textit{Near Surface Geophysics}, \textbf{7}, 447-462.

	\bibitem{Caboussat2012}
		Caboussat, A. \& Glowinski, R., 2012. 
		Regularization methods for the numerical solution of the divergence equation div u = f, 
		\textit{J. Comput. Math.}, \textbf{30}(4), 354-380.

	\bibitem{Castermant2008}
		Castermant, J., Mendon\c{c}a, C.A., Revil, A., Trolard, F., Bourri{\'e}, G.  \& Linde, N., 2008
		Redox potential distribution inferred from self-potential measurements associated with the corrosion of a burden metallic body,
		\textit{Geophys. Prospecting}, \textbf{56},269-282.

	\bibitem{Clement1993}
		Cl{\`e}ment, P. \& Li, S., 1993,
		Abstract parabolic quasilinear equations and application to a groundwater flow problem,
		\textit{Adv. Math.Sci. Appl.}, \textbf{3},17-32.

	\bibitem{ElBadia1998}
		El Badia, A. \& Ha-Duong, T. 1998. 
		Some remarks on the problem of source identification from boundary measurements, 
		\textit{Inverse Problems}, \textbf{14}, 883-891.

	\bibitem{ElBadia2000}
		El Badia, A. \& Ha-Duong, T. 2000. 
		An inverse source problem in potential analysis, 
		\textit{Inverse Problems}, \textbf{16}, 651-663.

	\bibitem{Geissert2006}  
		Gei{\ss}ert, M., Heck, H. \& Hieber, M., 2006,
		On the Equation div u = g and Bogovskii's Operator in Sobolev Spaces of Negative Order,
		in \textit{Partial Differential Equations and Functional Analysis},
		Birkh{\"a}user Basel, Basel.

	\bibitem{Glowinski2003}
		Glowinski, R., 2003. 
		Finite element methods for incompressible viscous flow, 
		\textit{Elsevier}.

	\bibitem{Guarracino2018}
		Guarracino, L., \& Jougnot, D., 2018. 
		A physically based analytical model to describe effective excess charge for streaming potential generation in water saturated porous media. 
		\textit{J. Geophys. Res.}, \textbf{123}, 52–65.

	\bibitem{Ikard2014}
		Ikard, S. J., Revil, A., Schmutz,M.,Karaoulis, M., Jardani,A. \& Mooney,M., 2014
		Characterization of focused seepage through an earthfill dam using geoelectrical methods,
		\textit{Groundwater}, \textbf{52}, 952-965. 

	\bibitem{Isakov}
		Isakov, V., 1990. Inverse source problems, in \textit{Mathematical surveys and monograph, 34},
		American Mathematical Society, Providence, Rhone Island.
	  
	\bibitem{Jardani2008}
		Jardani, A., Revil, A., Bol{\`e}ve,A., \& Dupont, J.P., 2008
		Three-dimensional inversion of self-potential data used to constrain the pattern of groundwater flow in geothermal fields,
		\textit{J. Geophys. Res.}, \textbf{113},B09204.

	\bibitem{Ling2005}
		Ling, L., Hon, Y.C. \& Yamamoto, M. 2005. 
		Inverse source identification for Poisson equation, 
		\textit{Inverse Problems Sci. Eng.}, \textbf{13}(4), 433-447.

	\bibitem{Magnoli1997}
		Magnoli, N. \& Viano, G. A. 1997. 
		The source identification problem in electromagnetic theory, 
		\textit{Journal of Mathematical Physics}, \textbf{39}(5), 2366-2338.

	\bibitem{Majeed2017}
		Majeed, M.U. \& Laleg-Kirati, T.M., M. 2017. 
		Iterative observer based method for source localization problem for Poisson in 3D, 
		\textit{American Control Conference}.


	\bibitem{Minsley2007}
		Minsley, B. J., Sogade, J., \& Morgan, F. D., 2007
		Three-dimensional source inversion of self-potential data, 
		\textit{Journal of Geophysical Research }, \textbf{112}, B2202.

	\bibitem{Nara2003}
		Nara, T. \& Ando, S., M. 2003. 
		A projective method for an inverse source problem of the Poisson equation, 
		\textit{IOP Inverse problems}, \textbf{19}, 355-369.

	\bibitem{Portniaguine2002}
		Portniaguine, O., Zhdanov, M.S., 2002
		3-D magnetic inversion with data compression and image focusing,
		\textit{Geophysics}, \textbf{67}(5), P.1532-1541.

	\bibitem{Prilepko}
		Prilepko, A.I., Orlovsky, D.G. \& Vasin, I.A.  2000.
		\textit{Methods for solving inverse problems in mathematical physics}, 
		Marcel Dekker, Inc., New York, Basel.

	\bibitem{Revil2013}
		Revil, A.\& Jardani, A., 2013. 
		The Self-Potential Method: Theory and Applications in Environmental Geosciences, 
		Cambridge: Cambridge University Press.

	\bibitem{Rittgers2013}
		Rittgers, J. B., Revil, A., Karaoulis, M., Mooney, M. A., Slater, L. D.,  \& Atekwana, E. A., 2013. 
		Self-potential signals generated by the corrosion of buried metallic objects with application to contaminant plumes, 
		\textit{Geophysics }, \textbf{78(5)}, EN65-EN82.

	\bibitem{Titov2015}
		Titov, K., Konosavsky, P. \& Narbut, M., 2015
		Pumping test in layered aquifer: numerical analysis of self-potential signals,
		\textit{J. Appl. Geophys.},\textbf{123},188-193.

	\bibitem{TrujilloBarreto2004}
		Trujillo-Barreto, N. J., Aubert-V{\'a}zquez, E., \& Vald{\'e}s-Sosa, P. A., 2004
		Bayesian model averaging in EEG/MEG imaging, 
		\textit{NeuroImage }, \textbf{21}, 1300-1319.

	\bibitem{Veen1997}
		Van Veen, B. D., Van Drongelen, W., Yuchtman, M. \& Suzuki, A., 1997
		Localization of brain electrical activity via linearly constrained minimum variance spatial filtering,
		\textit{IEEE Transactions on Biomedical Engineering},\textbf{44}, 867-880.

	\bibitem{Wesseling2001}
		Wesseling, P. 2001. 
		Principles of Computational Fluid Dynamics, 
		in \textit{Springer Series in Computational Mathematics, vol. 29},
		Springer.

	\bibitem{Yamatani1998}
		Yamatani, K. \& Ohnaka, K., 1998
		A reliable estimation method of a dipole for three-dimensional Poisson equation,
		\textit{J. Comput. Appl. Math.},\textbf{95}(1–2),139-151 .

\end{thebibliography}
\end{document}